\documentclass[preprint,letterpaper,amsmath,amssymb,floatfix,nofootinbib]{revtex4}

\usepackage{graphicx,epsfig}
\usepackage{amsmath,amssymb}
\usepackage{epsfig}
\usepackage{subfigure}
\usepackage{enumerate}
\usepackage{verbatim}

\newcommand{\be}{\begin{equation}}
\newcommand{\ee}{\end{equation}}
\newcommand{\ud}{\mathrm{d}}

\begin{document}

\title{Axions, Inflation and the Anthropic Principle} \author{Katherine J. Mack}\email{mack@ast.cam.ac.uk}
\affiliation{Kavli Institute for Cosmology, Institute of Astronomy, University of Cambridge, Madingley Road, Cambridge, CB3 0HA, UK}
\affiliation{Department of Astrophysical Sciences, Peyton Hall - Ivy Lane,  Princeton, NJ 08544, USA}


\begin{abstract}

The QCD axion is the leading solution to the strong-CP problem, a dark matter candidate, and a possible result of string theory compactifications.  However, for axions produced before inflation, symmetry-breaking scales of $f_a \gtrsim 10^{12}$ GeV (which are favored in string-theoretic axion models) are ruled out by cosmological constraints unless both the axion misalignment angle $\theta_0$ and the inflationary Hubble scale $H_I$ are extremely fine-tuned.  We show that attempting to accommodate a high-$f_a$ axion in inflationary cosmology leads to a fine-tuning problem that is worse than the strong-CP problem the axion was originally invented to solve.  We also show that this problem remains unresolved by anthropic selection arguments commonly applied to the high-$f_a$ axion scenario.

\end{abstract}

\maketitle

\section{Introduction} \label{s:Introduction}

The axion lies at the intersection of particle physics, cosmology and string theory, potentially playing a crucial role in each.  If it exists, the axion solves the strong-CP problem of quantum chromodynamics (QCD), fits easily into a string theoretic framework, and appears cosmologically as a form of cold dark matter.  However, the axion scenario faces a challenge as astrophysical and cosmological observations place increasingly tight constraints on axion and inflationary parameters.  

The axion solution to the strong-CP problem \cite{tHooft:1976a,tHooft:1976b} involves the introduction of a new global $U(1)$ symmetry in the early universe, known as the Peccei-Quinn (PQ) symmetry \cite{Peccei:1977hh}, which is broken at the symmetry-breaking scale $f_a$.  The resulting pseudo-Nambu-Goldstone boson is the axion \cite{Wilczek:1978,Weinberg:1978}.  Through the effects of instantons, the axion field's potential develops a minimum at the CP-conserving point, thus naturally explaining the lack of observed CP violation in the strong interactions. 

The current constraints on the axion allow two problematic scenarios. (1) The {\it low-$f_a$} axion, with $f_a \lesssim 10^{12}$ GeV, produced after cosmological inflation, exists in a narrow parameter space difficult to achieve in a string theory model and requires tuning of the underlying unified theory to avoid production of cosmologically catastrophic domain walls. (2) The {\it high-$f_a$} axion, with $f_a \gtrsim 10^{12}$ GeV, produced prior to inflation, exists at $f_a$ scales more easily accessible to string theory models but requires very careful tuning of the inflationary Hubble scale $H_I$ and the axion's initial condition $\theta_0$ to evade cosmological constraints.  While some models have been designed to produce string-theoretic axions with $f_a \sim 10^{12}$ GeV \citep[e.g.,][]{Dasgupta:2008hb}, generic string theory models produce axions with $f_a \sim 10^{16}$ GeV, near the string scale \citep{Svrcek:2006yi}.  Some have suggested that in this case, tuning of $\theta_0$ may come naturally from anthropic considerations \citep{Linde:1991km,Wilczek:2004cr,Tegmark:2005dy,Hertzberg:2008wr,Freivogel:2008qc}.

Previous works considering a high-$f_a$ axion have addressed the relationship between $H_I$ and cosmological constraints on axions.  In \cite{Fox:2004kb}, it was pointed out that if $H_I \gtrsim 10^{13}$ GeV, which corresponds to an inflationary energy scale $E_I \gtrsim 10^{16}$ GeV, primordial gravitational waves would be produced at a magnitude that would be detectable by the Planck satellite through its measurement of the cosmic microwave background (CMB) tensor-scalar ratio $r$ \citep{Kamionkowski:2002mi,Cooray:2003da}.  Such high values of $H_I$ would also result in large inflationary fluctuations to the axion's misalignment angle, so even for very small values of $\theta_0$, the axion field would overproduce isocurvature fluctuations in the CMB.  The conclusion of \cite{Fox:2004kb} was that a detection of inflationary gravitational waves by Planck would confirm a high $H_I$ and therefore rule out a high-$f_a$ (string-theoretic) QCD axion.

A recent paper by Hertzberg, Tegmark and Wilczek \cite{Hertzberg:2008wr} approached this issue in a converse way, stating that the assumption of the existence of a high-$f_a$ axion predicts a {\it low} $H_I$ and therefore the {\it non}-detection of inflationary gravitational waves.  They argue that anthropic selection on the string landscape supports both the existence of a high-$f_a$ axion at a non-negligible abundance and the very low $\theta_0$ required to evade cosmological constraints.  

In a companion paper \citep{MackSteinhardt:2009}, we have looked at this issue as applied to multiple string-theoretic axion-like fields.  We pointed out that string theory models that include a QCD axion (generally, a high-$f_a$ axion) can also be expected to produce other axion-like fields at similar masses, and the cumulative impact of additional axion-like fields overproduces dark matter and/or isocurvature modes unless the axion model and $H_I$ are extremely fine-tuned.

In this paper, we argue that the high-$f_a$ axion scenario is too problematic to be considered an appealing solution, {\it whether or not} (a) inflationary gravitational waves are detected by near-term satellite missions, (b) anthropic selection determines the density of dark matter, or (c) the QCD axion arises out of string theory.  We show that since both $\theta_0$ and the inflationary Hubble scale must be restricted to extremely low values for the axion to evade cosmological constraints, a viable axion model requires incredibly careful fine tuning.  The axion's tuning problem is so extreme that it rivals or exceeds that of the strong-CP problem that the axion was originally invented to solve.  Anthropic selection arguments do not alleviate this tuning problem: the most stringent constraints on the axion come not from the (possibly anthropic) dark matter density, but from cosmological observables whose values are not anthropically selected.  While the tuning problem in QCD that inspired the axion's proposal was troubling enough to require revision of that otherwise very successful theory, a tuning problem of equal or worse magnitude in the axion itself -- a hypothetical particle for which there is no compelling observational evidence -- could negate the axion's purpose completely.

To quantify the level of fine tuning required for an axion model to evade cosmological constraints, we use a figure of merit \citep{MackSteinhardt:2009} defined as the product of the inflationary slow-roll parameter $\epsilon$ and the axion misalignment angle, $\theta_0$, which measures the angular distance of the field from the CP-conserving $\theta=0$ point at the onset of oscillation.  The angle $\theta_0$ is a stochastic initial condition uniformly distributed in the interval $[0,\pi]$.  With this definition, ${\cal F}$ measures the volume of parameter space one is restricted to by observational constraints, taking into account tuning of both the initial condition of the axion and the level of tuning of the inflationary model.  An axion with $\theta_0\sim \mathcal{O}(1)$ existing in a simple, minimally tuned single-field inflationary model has ${\cal F} \sim 10^{-2}$.  We show in \S \ref{s:tuning} that cosmological constraints strongly rule out such a model, and that for the parameter space considered, the best-case scenario still allowed by constraints has ${\cal F} \sim 10^{-11}$.  This indicates that the tuning problem caused by trying to accommodate the existence of a high-$f_a$ axion in inflationary cosmology at best rivals the tuning required to solve the strong-CP problem without an axion [$\mathcal{O}(10^{-10})$], and, in most cases, far exceeds it.

The paper is organized as follows.
In \S \ref{s:constraints}, we discuss the production of the high-$f_a$ QCD axion and the origin of its cosmological constraints.  In \S \ref{s:tuning} we discuss the high degree of fine tuning required for high-$f_a$ QCD axion models to remain consistent with inflationary cosmology and cosmological constraints.  In \S \ref{s:Bayes}, we show how Bayesian model comparison disfavors the high-$f_a$ axion model in the context of inflation.  In \S \ref{s:anthropic}, we explain how anthropic arguments have been used to justify the choice of initial conditions for the axion field, and in \S \ref{s:anthropicfail} we show that anthropic selection arguments are insufficient to excuse the degree of fine tuning required for the axion to evade cosmological constraints.  We then discuss the implications of these results in \S \ref{s:discussion}.

\section{Cosmological constraints}
\label{s:constraints}

The major cosmological constraints on axions come from the measured density of dark matter and from the ratio of isocurvature to adiabatic fluctuations in the CMB.  We briefly review here the connection between axion properties and observables; for a more complete description of the origin of these constraints, we refer the reader to, e.g., \cite{Fox:2004kb,Hertzberg:2008wr}.

Axion particles are produced by the oscillation of a scalar field in a cosine-like potential after the breaking of a $U(1)$ symmetry known as the PQ symmetry to an $N$-fold degenerate vacuum \citep{Peccei:1977hh}.  The shape of the axion potential is a function of the field value $a \equiv f_a \theta$:
\be
V(a) = m_a^2 f_a^2 \left( 1 - \cos{[a/f_a]} \right).
\ee
The amplitude and duration of the oscillations determines the number of axion particles produced, and thus the axion contribution to the dark matter density.  After inflation, the field is uniform within the horizon volume at an angular position $\theta_0$, known as the misalignment angle, and oscillates about the minimum at $\theta=0$.  The oscillation is overdamped by Hubble expansion until $3H \simeq m_a(T_{osc})$, where $H$ is the Hubble scale, $m_a(T)$ is the axion mass, and $T_{osc}$ is the temperature at which oscillation begins.  The temperature dependence of $m_a(T)$ leads to two cases for the effects of axion oscillation depending on whether oscillation begins before or after QCD becomes strongly coupled at $T \sim 200$ MeV.  We distinguish the cases by the symmetry-breaking scale in comparison to $\hat{f}_a$, which is defined by
$\hat{f}_a \sim 0.26 (\Lambda/ 200 \textrm{ MeV})^2 m_{Pl}$, where $m_{Pl} \approx 2.4 \times 10^{18}$ GeV is the reduced Planck mass.
In the case of $f_a \lesssim \hat{f}_a$, the oscillation begins before $T \sim 200$ MeV and the mass of the axion has not yet reached a stable value at the onset of oscillation.  For higher values with $f_a \gtrsim \hat{f}_a$, oscillation starts later ($T_{osc} \lesssim 200$ MeV) and the axion mass may be considered to be constant during oscillation.
The magnitude of oscillations is determined by a combination of $f_a$, $\theta_0$, and the perturbations to $\theta_0$ from inflationary fluctuations, which have a mean square fluctuation $\sigma_\theta^2 \equiv \langle (\theta- \theta_0 )^2 \rangle $.  This is related to the Hubble parameter at inflation $H_I$ by \citep{Fox:2004kb}
\be \label{eq:sigma}
\sigma_\theta \approx \frac{H_I}{2 \pi f_a}.
\ee

When the axion field oscillates, the resulting axion particles form a cold Bose-Einstein condensate with a number density
\be
n_a \propto \zeta(T_{osc}) m_a f_a^2 (\theta_0^2 + \sigma_\theta^2)
\ee
where $m_a=m_a(0)$ and $\zeta(T) \equiv m_a(T) / m_a$ accounts for the temperature dependence of the axion mass \citep{Fox:2004kb}.  We can convert this number density to a temperature-independent quantity, the ratio $\xi_a$ of the axion energy per photon (in energy units), defined by \citep{Hertzberg:2008wr}
\be
\xi_a \equiv \frac{\rho_a (T_0)}{n_\gamma (T_0)} = \frac{m_a (T_0)}{m_a (T_{osc})} \frac{\rho_a(T_{osc})}{n_\gamma (T_0)} \frac{s (T_0)}{s (T_{osc})}
\ee
where $\rho_a$ is the axion density, $n_\gamma$ is the photon number density and $s$ is the entropy density.  In terms of the axion parameters, this becomes
\be
\xi_a \approx \Lambda (\theta_0^2 + \sigma_\theta^2) G .
\ee
The function $G$ accounts for the two cases of axion mass temperature dependence:
\be
G \approx
\begin{cases} 
  2.8 \left( \frac{\Lambda}{200 \textrm{ MeV}} \right)^{2/3} \left( \frac{f_a}{m_{\textrm{Pl}}} \right)^{7/6}  & \mbox{for }f_a \lesssim \hat{f}_a \\
  4.4 \left( \frac{f_a}{m_{\textrm{Pl}}} \right)^{3/2} & \mbox{for }f_a \gtrsim \hat{f}_a.
\end{cases}
\ee
Recently, new calculations of the temperature dependence of the axion mass have been published \cite{Wantz:2009it} that slightly modify this function, but the change is not significant for our purpose.
In order to set a constraint on the axion parameters based on the dark matter density, we compare $\xi_a$ to the ratio of the dark matter energy per photon, $\xi_{CDM} \approx 2.9$ eV.

Fluctuations in the axion field during inflation produce isocurvature-type perturbations that can be observed in the CMB.  Limits on the fraction of CMB perturbations that are isocurvature can therefore place constraints on an axion field existing during inflation \citep{Fox:2004kb,Hamann:2009yf}.  To test against this limit, we calculate the ratio of the average power in the isocurvature component to the average total power in CMB temperature fluctuations:
\be
\alpha_a \equiv \frac{\langle (\delta T/T)^2_{iso} \rangle}{\langle (\delta T/T)^2_{tot} \rangle}.
\ee
Current constraints from CMB observations limit $\alpha_a < 0.072$ \citep{Komatsu:2008}.  In terms of axion parameters, the ratio is
\be
\alpha_a \approx \frac{8}{25} \frac{(\Lambda/\xi_m)^2}{\langle (\delta T/T)^2_{tot}) \rangle} \sigma_\theta^2 (2 \theta_0^2 + \sigma_\theta^2) G^2
\ee
where $\xi_m = 3.5$ eV is the matter density per photon and $\langle (\delta T/T)^2_{tot} \rangle \approx (1.1 \times 10^{-5})^2$.

Note that in the case of isocurvature modes, the dependence on $H_I$ (via $\sigma_\theta$) and the axion symmetry breaking scale $f_a$ are much stronger than in the case of dark matter density.  The strong dependence on $\sigma_\theta$ leads to the isocurvature constraint primarily ruling out high values of $H_I$.  The behavior of the dark matter density constraint depends primarily upon the relationship between $\theta_0$ and $\sigma_\theta$.  When $\theta_0 << \sigma_\theta$, the oscillation of the field is dominated by inflationary fluctuations and the density constraint provides a limit on $H_I$.  When $\theta_0 >> \sigma_\theta$, however, the constraint mainly rules out high values of $f_a$.

\begin{figure}[h!]
\centering
\subfigure[]{%
\epsfig{figure=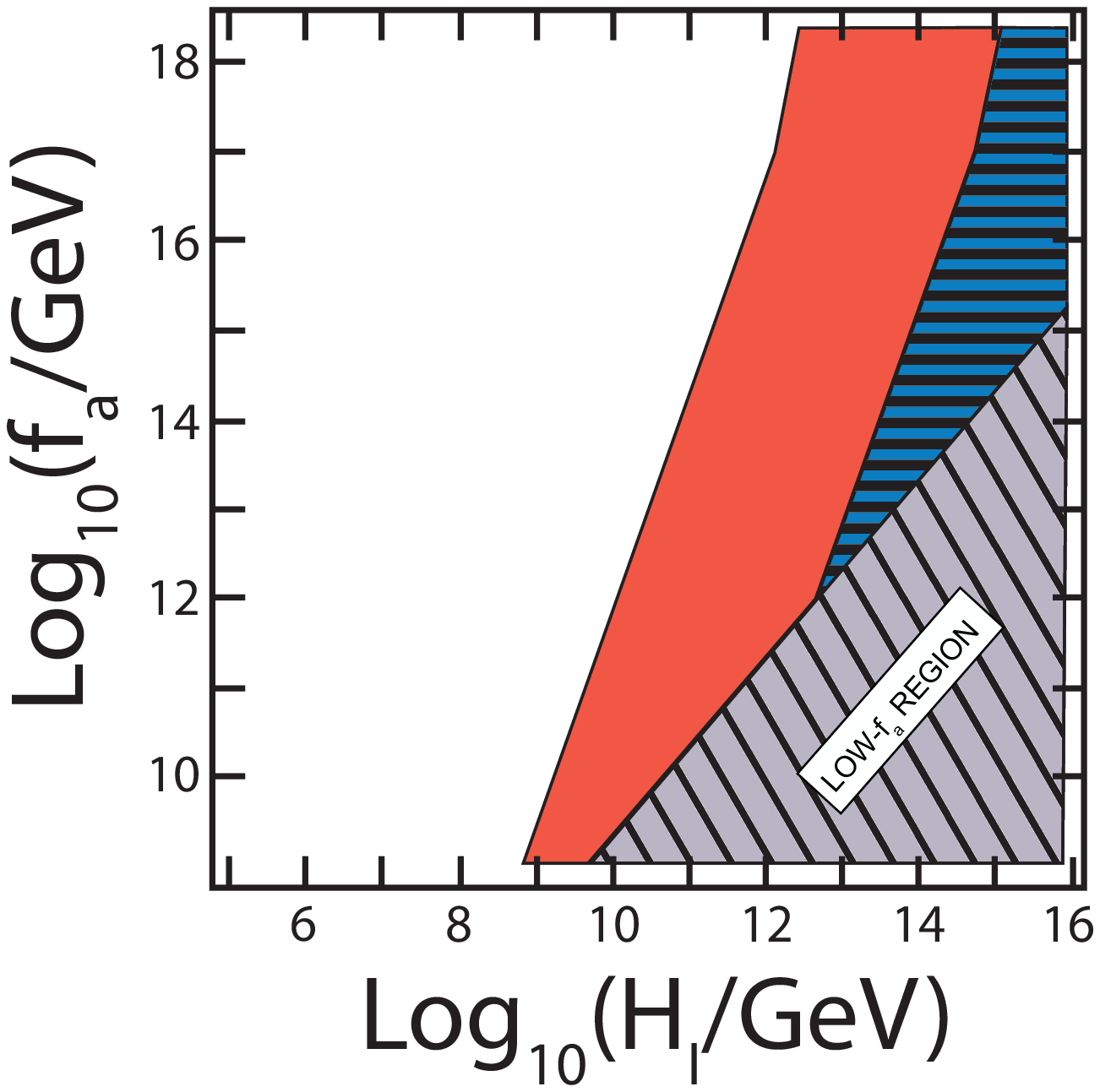,width=2in,angle=0}} \quad
\subfigure[]{%
\epsfig{figure=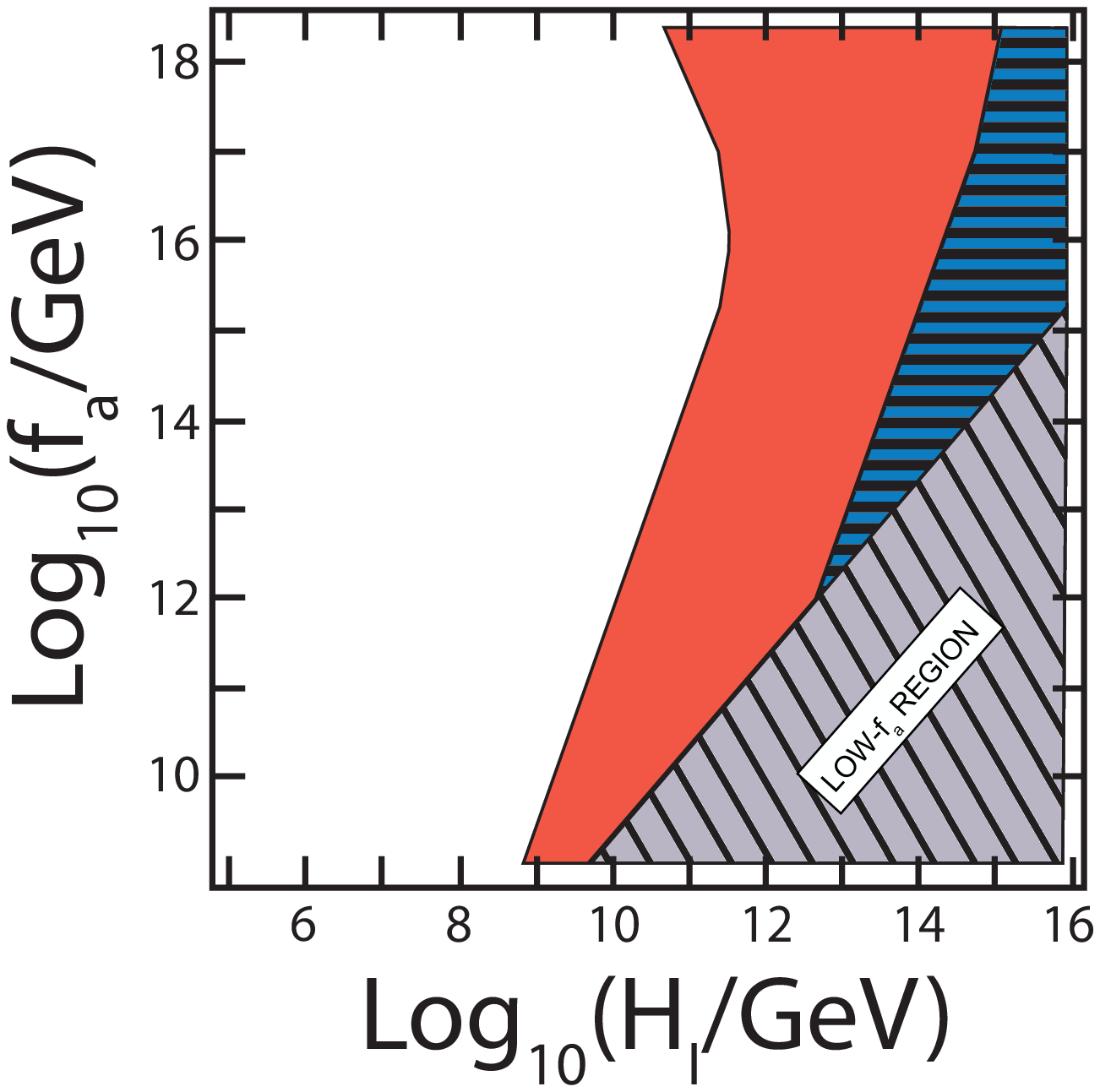,width=2in,angle=0}} \\
\subfigure[]{%
\epsfig{figure=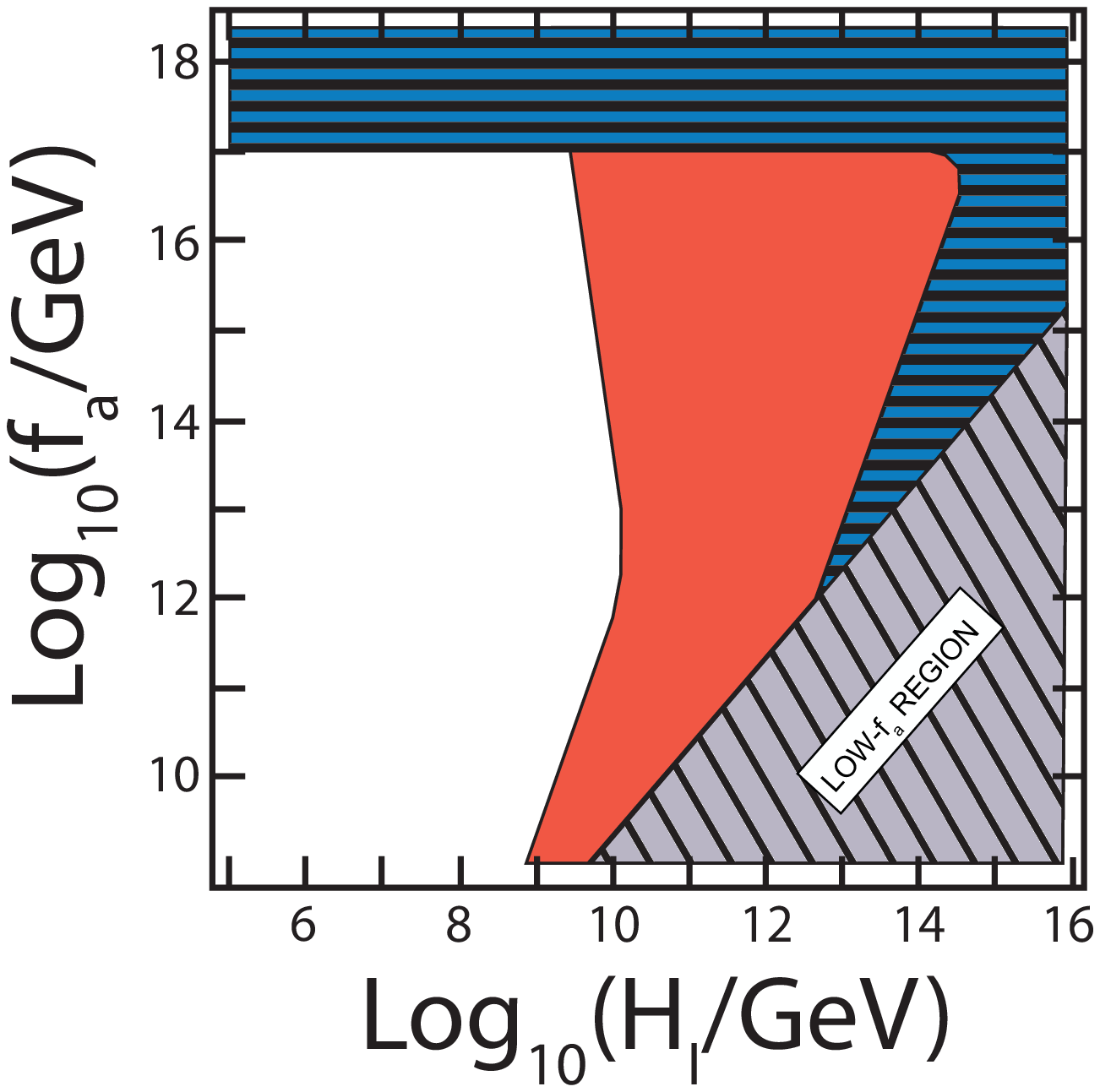,width=2in,angle=0}} \quad
\subfigure[]{%
\epsfig{figure=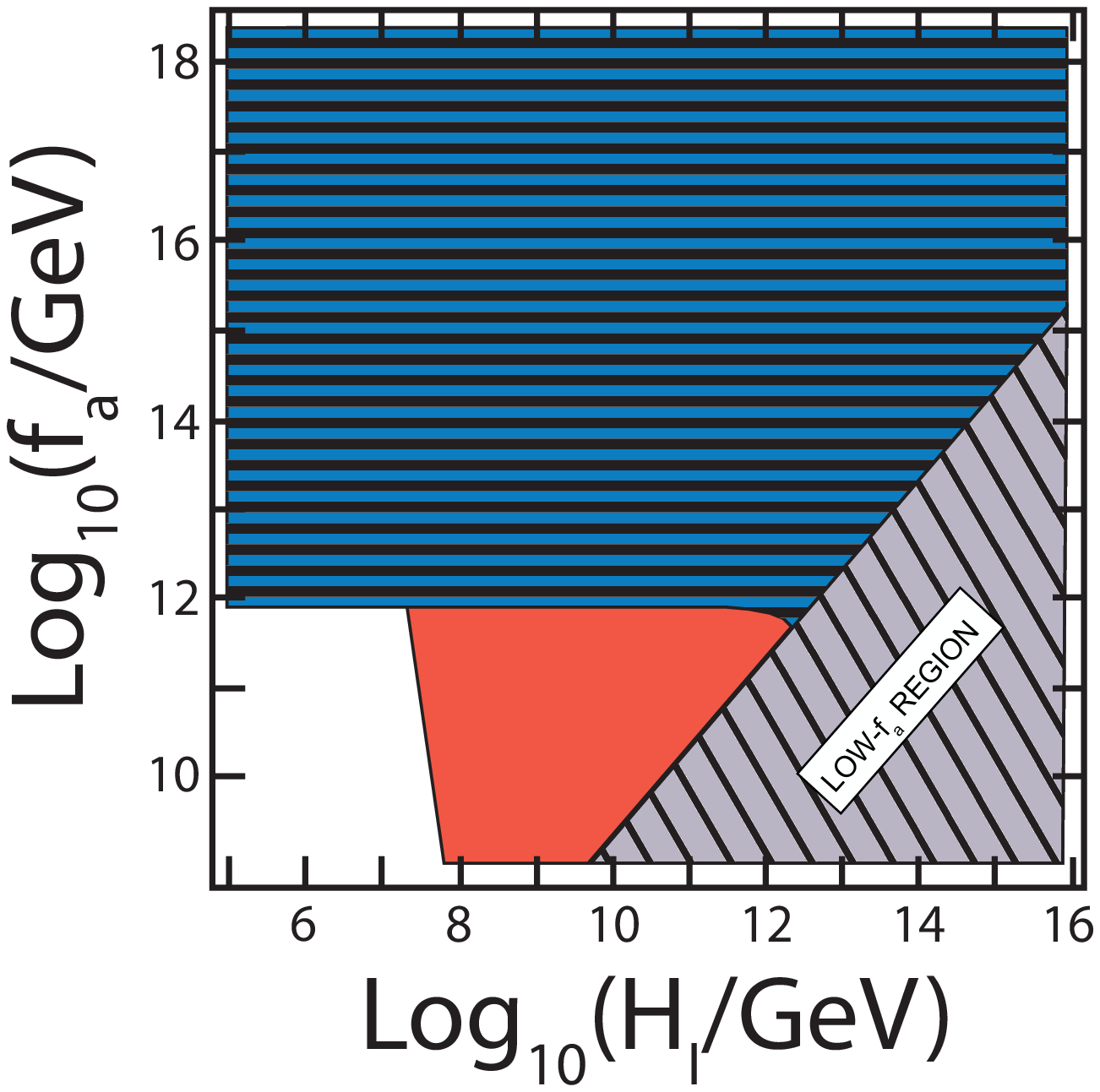,width=2in,angle=0}}
\caption{Dark matter density (blue, horizontally hatched) plotted over CMB isocurvature (red) constraints for $\theta_0=$ (a) $10^{-10}$, (b) $10^{-5}$, (c) $10^{-3}$, (d) $1$.  The x-axis ($H_I$) is the Hubble scale during inflation and the y-axis ($f_a$) is the axion symmetry-breaking scale.  The grey diagonally hatched region in the lower right represents the parameter space of low-$f_a$ axions whose symmetry breaking occurs after inflation.  White regions are unconstrained for the given value of $\theta_0$.}%
\label{f:red}%
\end{figure}

In Figure \ref{f:red}, we show how the density constraint and the isocurvature constraint change as $\theta_0$ is increased from $10^{-10}$ to $1$.  The parameter space we plot is ($H_I,f_a$).  For four values of $\theta_0$, we plot the density constraint (blue, with horizontal hatching) over the isocurvature constraint (red, unhatched).  The low-$f_a$ axion regime is the grey diagonally hatched region in the lower right corner of each plot; we do not consider the low-$f_a$ regime in this work.  Regions that remain white are allowed within the cosmological constraints for the stated value of $\theta_0$.  From these plots, we see that at the lowest values of $\theta_0$ [panels (a) and (b)], both constraints primarily limit $H_I$, but the isocurvature constraint is much more restrictive.  As $\theta_0$ is increased toward more natural values [i.e., values closer to $\mathcal{O}(1)$; panels (c) and (d)], the isocurvature constraint remains more restrictive against high values of $H_I$, but the density constraint begins to rule out high values of $f_a$.  For $\theta_0 = 1$ [panel (d)], the only region of the plot remaining unconstrained is for $f_a \lesssim 10^{12}$ GeV and $H_I \lesssim 10^7$ GeV; a combination of both the isocurvature and density constraints rules out the rest of the area in the plot.

A recent work \citep{Arvanitaki:2009fg} has suggested a new constraint on the QCD axion from the non-detection of black hole superradiance. This limit would restrict the axion parameter space further to $f_a \lesssim 2 \times 10^{17}$ GeV.


\section{Fine-tuning of axion and inflationary parameters}
\label{s:tuning}

The production of high-$f_a$ axions and of the resulting isocurvature modes depends on the PQ symmetry-breaking scale $f_a$, the axion field's misalignment angle $\theta_0$, and the Hubble scale at inflation, $H_I$.  The symmetry-breaking scale might eventually be predicted from a string theory model; in the most generic string theoretic axion production models, $f_a$ is close to the string scale, around $10^{16}$ GeV.  For these models to be viable, $\theta_0$ must be small in order to avoid constraints from the observed dark matter density and/or the CMB isocurvature fraction.  This angle is randomly selected by the field before inflation and is uniform throughout our horizon.  Therefore, we can classify any model with a small value of $\theta_0$ to be tuned to a degree proportional to the smallness of $\theta_0$.

The Hubble scale during inflation can also be considered a parameter of a high-$f_a$ axion scenario, since it determines the magnitude of inflationary perturbations ($\sigma_\theta$, defined in \S \ref{s:constraints}) to $\theta_0$.  Even in a case in which $\theta_0=0$ exactly, axions will still be produced due to the displacement of the field from the $\theta=0$ point by inflationary fluctuations.  

For simple, single-field inflation, there is a relationship between $H_I$ and the slow-roll parameter $\epsilon$, which measures the flatness of the inflationary potential and is defined by
\be
\epsilon = \frac{m_{\textrm{Pl}}^2}{2} \left(\frac{V'}{V}\right)^2.
\ee
This parameter is related to the inflationary equation of state $w$ through $\epsilon = \frac{3}{2} (1+w)$.  Since $\epsilon \approx 10^{10}(H_I/m_{Pl})^2$, values of $H_I \lesssim 10^{13}$ GeV correspond to values of $\epsilon <<1$ or equivalently $(1+w) << 1$ (i.e., the Universe in an almost exactly deSitter state).  For these models, the inflationary potential is flatter than necessary to sustain 60 e-folds of inflation and solve the horizon problem, and can therefore be considered fine-tuned.  Therefore, the level of tuning is quantifiable by the smallness of the slow-roll parameter $\epsilon$.

\begin{figure}[h!]
\centering
\epsfig{figure=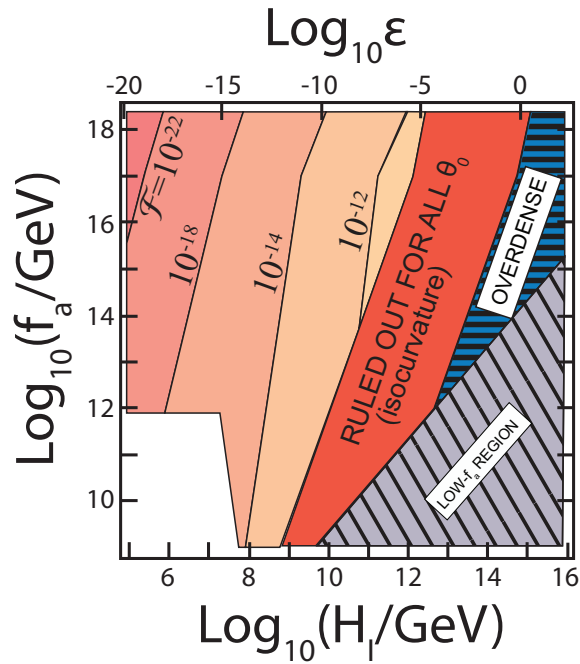,width=3in,angle=0}
\caption{Contours of fine-tuning figure of merit $\mathcal{F}$ [defined in equation (\ref{eq:F})] for regions of parameter space allowed by current cosmological constraints.  The isocurvature constraint for all $\theta_0$ is the red region labelled ``ruled out for all $\theta_0$ (isocurvature)'' and the dark matter density constraint for all $\theta_0$ is labelled ``overdense.''  The grey diagonally-hatched region labelled ``low-$f_a$ region'' is the part of parameter space in which the low-$f_a$ axion occurs after inflation.  The upper axis indicates the value of the slow-roll parameter $\epsilon$, discussed in \S \ref{s:tuning}.}%
\label{f:tuning}%
\end{figure}

In \S \ref{s:Introduction}, we briefly described our fine-tuning figure of merit, 
\be \label{eq:F}
\mathcal{F} \equiv \epsilon \times \theta_0,
\ee
which quantitatively accounts for the total tuning of a high-$f_a$ axion model.
In Figure \ref{f:tuning}, we show contours of the $\mathcal{F}$ values required in each part of the parameter space $(H_I,f_a)$ to evade existing cosmological constraints.  As in Figure \ref{f:red}, the diagonally hatched region in the lower right is the low-$f_a$ axion regime which we do not address here.  The blue horizontally hatched region labelled ``overdense'' is the part of the parameter space ruled out by the dark matter density for any value of $\theta_0$, and the red region labelled ``ruled out for all $\theta_0$ (isocurvature)'' is the region for which isocurvature constraints rule out axion models with any value of $\theta_0$.  The white region in the lower left is unconstrained for values of $\theta_0 \sim 1$; in that region, the value of $\mathcal{F}$ is determined by the slow-roll parameter $\epsilon$, which can be read off the top horizontal axis.  For the region plotted, the largest (least-tuned) value of $\mathcal{F}$ that can be achieved is $\lesssim 10^{-11}$, which we note is comparable to the order of magnitude of the strong-CP problem the axion was invented to solve.


\section{Bayesian model comparison}
\label{s:Bayes}

Another way to look at the tuning of the high-$f_a$ QCD axion model is to perform a Bayesian model comparison between a model in which axions exist and a model in which they do not.  The posterior probability of a model $M_i$ is given by
\be
P(M_i | D) = \frac{P(D|M_i) P(M_i)}{P(D)}
\ee
where $D$ is the data.  If the probability of an observation, given a physical model $M_i$, is known, this provides the probability $P(D|M_i)$.  $P(D)$ is the unconditional marginal likelihood of the data; it can be expressed as an integral over all models, $\int P(D|M')P(M')\ud M'$.  For two models, $M_0$ and $M_1$, the Bayes factor $B$ is then defined as the ratio of the posterior probabilities, $B_{M_1,M_0} = P(M_1 | D) / P(M_0 | D)$, where $M_1$ is the model with the larger marginal likelihood.  A Bayes factor between 3 and 20 is considered to be ``positive'' evidence against model $M_0$, whereas Bayes factors from 20 to 150 constitute ``strong'' evidence against and a factor above 150 is ``very strong'' evidence against model $M_0$ \cite{Kass:1995}.

In a companion paper \cite{MackSteinhardt:2009}, we have calculated the Bayes factors for scenarios in which axions, inflation, or string theory (which produces additional axion-like fields) are excluded from the model, and we have found that, given the cosmological observations, models that combine all three elements are exponentially less favored than models for which one or more is excluded.  In particular, we found that a comparison between (a) a model with a high-$f_a$ QCD axion in the context of inflation and string theory (which suggests $f_a \sim 10^{16}$ GeV), and (b), an otherwise identical model with no QCD axion (and an unsolved strong-CP problem), the Bayes factor was 
\be
B_{(\textrm{no axion,axion})} = \frac{P(\textrm{no axion})}{P(\textrm{axion})} \times 10^{12} \gtrsim 10^2 .
\ee
Here, $P(\textrm{no axion})/P(\textrm{axion}) \approx 10^{-10}$ accounts for the lack of a solution to the strong-CP problem in a model with no axions, and we have assumed for this example that there are no additional axion-like fields from string theory present.  This shows that a high-$f_a$ QCD axion is strongly disfavored even if no other axion-like fields are produced, and when the axion model is given the maximal theoretical prior weighting through the assumption that no other solution to the strong-CP problem can be found.

If it is taken as a given that a high-$f_a$ QCD axion exists, however, and instead we compare models with and without inflation, we see that generic inflationary models are exponentially disfavored.  Comparing a model with a generic form of inflation to one in which we have an alternative to inflation in which the axion is never excited from its minimum (e.g., the cyclic model \cite{Steinhardt:2006bf}), the Bayes factor is
\be
B_{(\textrm{no infl,infl})} = \frac{P(\textrm{no  infl})}{P(\textrm{infl})} \gtrsim \frac{P(\textrm{no  infl})}{P(\textrm{infl})} 10^{12} ,
\ee
where the ratio $P(\textrm{no  infl})/P(\textrm{infl})$ quantifies any theoretical priors that would favor or disfavor an inflationary solution compared with an alternative, and again we have assumed no additional fields.  We see that unless inflation is considered to be exponentially more plausible than an alternative model, when high-$f_a$ axions are assumed, the limit on the isocurvature mode extremely strongly disfavors standard inflation, whether or not additional string-theoretic axion-like fields are produced.

Our conclusion based on the Bayesian model comparison is that the fine tuning of the models required to make a high-$f_a$ QCD axion viable in the context of inflationary cosmology is so extreme as to challenge the compatibility of the two paradigms.  This conclusion is similar to that reached in \cite{MackSteinhardt:2009}, but in this case we do not base our assumptions on whether the QCD axion arises from string theory.  However, if we take into account the expectation that string-theoretic axions would have $f_a$ values in the vicinity of $10^{16}$ GeV (the string scale), a challenge to high-$f_a$ axions is by extension a challenge to any axions arising from string theory.

In the next section, we will discuss how anthropic arguments have been used in an attempt explain the tuning of the axion misalignment angle required to evade cosmological constraints.  The subsequent section discusses why anthropic arguments are ultimately insufficient to explain the tunings and do not alleviate the problem of reconciling axions with inflation.

\section{Anthropic explanation for initial conditions}
\label{s:anthropic}

For high-$f_a$ axions, the fact that inflation makes $\theta_0$ uniform within a horizon volume allows for the possibility that in our horizon, $\theta_0$ may be much smaller than the global average.  A very small $\theta_0$ would allow the axion to solve the strong-CP problem without producing dark matter at a higher density than observed.
This situation could come about either through a fortuitous accident of nature or through anthropic selection, if it is true that a high dark matter density hinders the development of observers such as ourselves.

In recent years, there has been a great deal of work focused on the latter scenario \citep[e.g.,][]{Hertzberg:2008wr,Linde:1991km,Tegmark:2004qd,Tegmark:2005dy,Freivogel:2008qc,Wilczek:2004cr,Arvanitaki:2009fg}.
According to these recent works, axions are ideal candidates for the application of anthropic selection because they represent a species in which the prior probability distribution $f_{prior}(\theta_0)$ for the relevant parameter ($\theta_0$) is well understood and the selection function on the landscape $f_{selec}(\theta_0)$ can be found based on calculations of the habitability of regions with different axionic dark matter densities.  The total anthropic probability distribution of the parameter $\theta_0$ is then the product of these two distributions, $f(\theta_0) = f_{prior}(\theta_0) f_{selec}(\theta_0)$.  For the axion, the prior probability distribution for $\theta_0$ is known to be uniform within the range $[0,\pi]$.  A selection function for the dark matter density, and by extension $\theta_0$, has been calculated numerically in \cite{Tegmark:2005dy} and also discussed in \cite{Wilczek:2004cr} and \cite{Hertzberg:2008wr}.  In essence, the argument is that a high dark matter density would result in overly dense galaxies, which would lead to an increase in the number of close stellar encounters experienced by a solar system.  These close encounters would render a solar system unstable; therefore life is less likely to evolve in regions of the string-theory landscape with high dark matter density (but see \cite{Hellerman:2005yi}).  To obtain their selection function, Tegmark {\it et al.}\cite{Tegmark:2005dy} first estimate the minimum and maximum densities of galaxy halos in which observers can form.  They convert this to a distribution of the fraction of protons in halos of different densities in terms of the density of dark energy, dark matter, and baryons, and the amplitude of scalar inflationary fluctuations.  They then marginalize this distribution over the dark energy density to obtain a probability distribution of the axion dark matter density.  This distribution is peaked at a value a factor of about 3 greater than the observed dark matter density, with a 95\% confidence interval containing values from about 2/3 to 100 times the observed dark matter density.

If such a selection function is used, the total probability distribution, taking into account the theoretical prior and the anthropic selection function, does not favor all values of $\theta_0$ equally.  Instead, the selection function skews the distribution in the direction of low $\theta_0$ values.  It is argued that the necessity for a low $\theta_0$ value to evade the dark matter density constraint should not be considered an arbitrary tuning of the model but rather a natural environmental effect, supported on the basis that we exist to observe the dark matter density at all \citep{Tegmark:2005dy}.

\section{Scope of the anthropic explanation}
\label{s:anthropicfail}

As mentioned above, the axion is often held up as an ``instructive example'' in which both the prior probability distribution and the selection effect can be taken into account and unambiguous predictions can be made \citep{Tegmark:2005dy}.  We have found, however, that while the selection effects hinge on the consequences of the overproduction of dark matter, the limits on the axion's properties are dominated by the {\it nonanthropic} CMB isocurvature fraction in much of the parameter space of interest.  Except in cases of very low-energy inflation (low $H_I$ values), the tuning of $\theta_0$ required to prevent the CMB isocurvature mode fraction from being larger than that observed is orders of magnitude greater than the tuning required to prevent the overproduction of dark matter.  If anthropic selection cannot account for the tuning, this tuning must be considered a challenge to the physical model.

An anthropic argument is considered compelling if the following criteria are met.  (1) The anthropic selection function (the probability for observers to develop at different values of the parameter in question) must be convincingly presented.  (2) The observed value must lie at a high-probability point in the selection function distribution.  However, both these criteria are problematic.  Criterion (1) requires possibly arbitrary assumptions about the definition of ``observer'' and marginalization over other parameters in the parameter space which might also affect the probability for the development of life.  Criterion (2) is difficult to evaluate, because it is unclear how a model might be disproved based on the distance of a parameter's value from the point of maximum probability on the landscape.  It has also been suggested \citep{Weinberg:1987dv,Hall:2007ja} that an anthropic explanation should be considered compelling if the observed parameter lies near the ``anthropic boundary;'' i.e., the observed value should not be tuned substantially beyond what anthropic selection requires.  Hall \& Nomura \cite{Hall:2007ja} describe an observation of a parameter near the anthropic boundary as evidence in favor of environmental selection.

\begin{figure}[h!]
\centering
\epsfig{figure=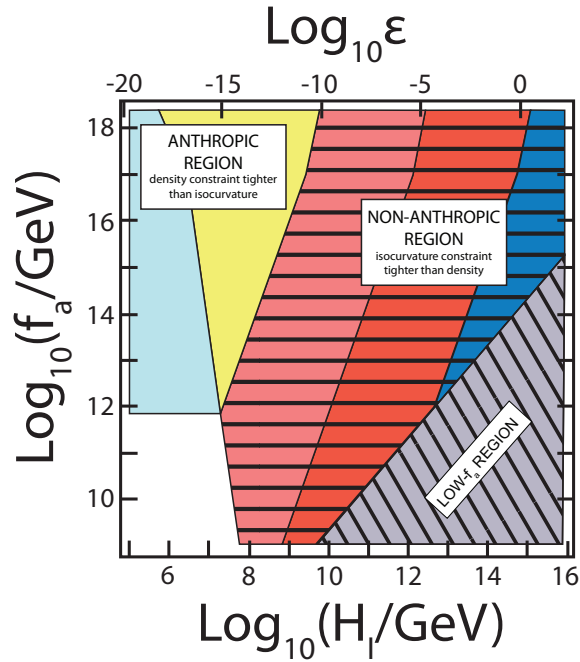,width=3in,angle=0}
\caption{Regions in the ($H_I,f_a$) parameter space for which dark matter density and CMB isocurvature fraction constraints dominate.  In the horizontally hatched ``non-anthropic region,'' the isocurvature fraction is a tighter constraint on $\theta_0$ than the dark matter density, and in the unhatched ``anthropic region'' the opposite is true.  From left to right, the colored regions are: (light blue) ruled out only by density; (yellow) ruled out first by density, then isocurvature; (pink) ruled out first by isocurvature, then density; (red) ruled out by isocurvature for all $\theta_0$ and (blue) ruled out by isocurvature and density for all $\theta_0$.  The grey diagonally-hatched region labelled ``low-$f_a$ region'' is the part of parameter space in which the low-$f_a$ axion occurs after inflation.  The upper axis indicates the value of the slow-roll parameter $\epsilon$, discussed in \S \ref{s:tuning}.}%
\label{f:QCDregimes}%
\end{figure}

In Figure \ref{f:QCDregimes}, we show the ($H_I,f_a$) parameter space divided into ``anthropic'' and ``non-anthropic'' regions.  In the horizontally hatched region, the tuning of $\theta_0$ required to evade the CMB isocurvature constraint is orders of magnitude more extreme than that required to avoid producing axionic dark matter at densities above the observed value.  We label this region ``non-anthropic'' to reflect the fact that anthropic selection, which relies on keeping the dark matter density within a range of livable values \citep{Tegmark:2005dy}, does not support the level of tuning needed to evade the observed constraint on the CMB isocurvature fraction.  Models in this region are not near the anthropic boundary \citep{Weinberg:1987dv,Hall:2007ja}.  In the unhatched region, the constraints on $\theta_0$ from the dark matter density are stronger than those from CMB isocurvature modes.  We label this region ``anthropic'' because an argument for a low $\theta_0$ based purely on avoiding the overproduction of dark matter is applicable there.  Note, however, that admitting an anthropic explanation over this region is still a charitable assumption, as the selection function does not strictly prohibit dark matter at levels higher than observed, and in fact allows a range that includes densities well in excess of the current constraint.

A fair comparison of the ``anthropic'' and ``non-anthropic'' regions must also take into account the necessary fine tuning of $H_I$ required in each part of the parameter space.  Figure \ref{f:QCDregimes} shows that the ``anthropic'' region appears only at values of $H_I \lesssim 10^{10}$ GeV; i.e., only in highly tuned inflation models ($\epsilon \lesssim 10^{-10}$).

We saw in Figure \ref{f:tuning} that before considering anthropic arguments, our measure of fine tuning $\mathcal{F}$ was at best $\mathcal{O}(10^{-11})$ for a small region of the parameter space considered, and many orders of magnitude worse over the bulk of the region plotted.  This indicates that in order to evade cosmological constraints, a successful axion model is forced to reside within a fraction $10^{-11}$ of the possible ($\theta_0,\epsilon$) parameter space. 

Now we ask if the application of the anthropic principle to this scenario is sufficient to explain the required tuning.  To make a conservative assumption of the tuning in the ``anthropic'' region after anthropic selection is taken into account, we use the assumption that there is a strong anthropic selection against existing in regions of the Multiverse with values of the dark matter density larger than we presently observe.  For a minimal estimate of the tuning in this region, therefore, we can ignore the unnatural smallness of $\theta_0$ and only factor in tuning of the inflationary model.  In this case, the necessary level of tuning is $\mathcal{F} = \epsilon$, which can be read off the $\epsilon$ axis along the top of the plot.  We see that for the highest values of $f_a$ plotted ($f_a \sim m_{Pl}$), the tuning required is $\lesssim 10^{-10}$, comparable to that of the strong-CP problem.  For values of $f_a \sim 10^{16}$ GeV (near the string scale), $\epsilon \lesssim 10^{-12}$, a tuning about two orders of magnitude worse than the strong-CP problem.

\section{Discussion}
\label{s:discussion}

We have shown that the fine tuning of axion and inflationary parameters required to accommodate a high-$f_a$ QCD axion in the context of a simple inflationary model is at best (in a small part of parameter space) comparable to the tuning that defines the strong-CP problem and much worse over the bulk of the parameter space considered.  In \S \ref{s:Bayes}, we also showed that a Bayesian model comparison disfavors the axion solution.  In \S\S \ref{s:anthropic} and \ref{s:anthropicfail}, we examined the possibility that anthropic selection effects could account for the apparent tuning of $\theta_0$.  In order to be considered successful in this context, we require that:
\begin{enumerate}
 \item Anthropic selection must provide a compelling argument for the smallness of the parameter ($\theta_0$, via the dark matter density), and,  \label{it:compelling}
 \item The anthropic observable (dark matter density) must provide a stronger limitation on the parameter than non-anthropic observables (in this case, the CMB isocurvature mode fraction). \label{it:dominant}
\end{enumerate}
Item \ref{it:dominant} must be satisfied if we live on the anthropic boundary, which some authors have suggested \citep{Weinberg:1987dv,Hall:2007ja}.  We assume for this purpose that Item \ref{it:compelling} is satisfied, but we have shown in \S \ref{s:anthropicfail} that Item \ref{it:dominant} is not satisfied in general.

In regions of the parameter space for which Item \ref{it:dominant} does hold, we re-examine the question of the fine tuning of the model, neglecting the (anthropically selected) smallness of $\theta_0$.  We find that fine tuning of $H_I$ is necessary for models in that region to evade cosmological constraints.  For $f_a \sim m_{Pl}$, the least-tuned inflationary models compatible with axions and cosmology have $\epsilon \lesssim 10^{-10}$, whereas for values of $f_a \sim 10^{16}$ GeV (as predicted in the most generic string theory models), $\epsilon \lesssim 10^{-12}$.
This would indicate that the axion solution to the strong-CP problem merely exacerbates the problem by moving it from particle physics to cosmology, all while requiring the assumption of an (as-yet) undetected particle.

We consider this to be a strong challenge to the high-$f_a$ axion scenario in the context of inflationary cosmology.  Since string-theoretic QCD axions generically appear at high $f_a$ scales, this is, by extension, a challenge to the realization of a QCD axion in string theory.  This challenge is made even stronger by the expectation of additional axion-like fields from string theory, as discussed in a companion paper \cite{MackSteinhardt:2009}.
We suggest that one of the following modifications of the axion scenario is required.
\begin{enumerate}[A.]
 \item One of the components of this scenario -- generic models of inflation, string theory (in fact any theory that requires a high $f_a$ scale), or axions -- must be abandoned.  \label{it:choosetwo}
 \item An anthropic argument against CMB isocurvature fractions higher than those observed, or against large inflationary energy scales, must be presented. \label{it:anthrowin}
\end{enumerate}

Option \ref{it:choosetwo} might be realized in a number of different ways.  If the axion solution to the strong-CP problem is to be preserved, one could abandon the attempt to find a high-$f_a$ QCD axion and instead assume a low-$f_a$ QCD axion model (with $N=1$ to prevent the formation of domain walls), for which a non-generic string-theoretic model could potentially be found (such as has been discussed for warped throats \citep{Dasgupta:2008hb}). 

Alternatively, high-$f_a$ QCD axions are consistent with early-universe models in which inflation does not strongly excite the axion field.  Hybrid inflation, in which inflation is driven by evolution of two scalar fields, can result in very low energy scales of inflation, with $H_I$ values as low as $\sim 10^4$ GeV \citep{Linde:1993cn}.  Any inflationary model with $H_I \lesssim 10^6$ GeV is unconstrained by isocurvature modes for $f_a \lesssim m_{Pl}$, and would require tuning of only $\theta_0$ to evade the dark matter density constraint.  In the cyclic model \citep{Steinhardt:2004gk}, light fields are not excited and therefore isocurvature modes and an excess of axionic dark matter would not be produced.  Failing to detect signatures of primordial gravitational waves with satellite missions such as Planck \citep{Fox:2004kb} would lend support to this option.

Finding a new solution to the strong-CP problem that does not involve an axion is another possibility, provided the axion-like fields naturally produced in string theory models do not themselves cause cosmological problems \citep{MackSteinhardt:2009}.  If anthropic selection arguments can be made against CMB isocurvature modes or high inflationary energy scales (Option \ref{it:anthrowin}), this could constitute another solution to the cosmological axion problem.  

In the meantime, we argue that the scenario we present here -- a QCD axion that exists during single-field inflation -- does not constitute a consistent model without some degree of modification.  As stated in the Introduction, this conclusion holds {\it even if}: $H_I$ is not high enough for near-term detectors to observe primordial gravitational waves, the dark matter density is indeed an anthropic parameter, and/or string theory is not the origin of the QCD axion.

Fine-tuning problems have long been considered ample motivation for the revision of a theory. In fact, the axion was hypothesized not for any observational or fundamental theoretical reason, but simply to solve a fine-tuning problem in QCD. If a theory's sole motivation is the solution of a fine-tuning problem, but it then produces an even more extreme tuning problem and an unseen particle, it is not a good theory. Unless an alteration of the axion model or early universe theory that does away with the tunings is adopted, the QCD axion should be considered an unhelpful complication of particle physics and an alternative solution to the strong-CP problem should be sought.

\section{Acknowledgments}
We wish to thank Paul Steinhardt for his comments and suggestions throughout the course of this work.  We also thank Viviana Acquaviva, Jacob Bourjaily, Fiona Burnell, Adam Burrows, Phil Chang, Joseph Conlon, Joanna Dunkley, Daniel Grin, Chris Hirata, David Spergel, David Spiegel and Daniel Wesley for helpful conversations.


\begin{thebibliography}{10}

\bibitem{tHooft:1976a}
 G. 't Hooft, Phys.\ Rev.\ Lett. {\bf 37}, 8 (1976).

\bibitem{tHooft:1976b}
 G. 't Hooft, Phys.\ Rev.\ D. {\bf 14}, 3432 (1976).

\bibitem{Peccei:1977hh}
  R.~D.~Peccei and H.~R.~Quinn,
  Phys.\ Rev.\ Lett.\  {\bf 38}, 1440 (1977).

\bibitem{Wilczek:1978}
  F. Wilczek, Phys.\ Rev.\ Lett.\  {\bf 40}, 279 (1978).

\bibitem{Weinberg:1978}
  S. Weinberg, Phys.\ Rev.\ Lett.\  {\bf 40}, 223 (1978).

\bibitem{Dasgupta:2008hb}
  K.~Dasgupta, H.~Firouzjahi and R.~Gwyn,
  JHEP {\bf 0806}, 056 (2008)
  [arXiv:0803.3828 [hep-th]].

\bibitem{Svrcek:2006yi}
  P.~Svrcek and E.~Witten,
  JHEP {\bf 0606}, 051 (2006)
  [arXiv:hep-th/0605206].

\bibitem{Hertzberg:2008wr}
  M.~P.~Hertzberg, M.~Tegmark and F.~Wilczek,
  Phys.\ Rev.\  D {\bf 78}, 083507 (2008)
  [arXiv:0807.1726 [astro-ph]].

\bibitem{Linde:1991km}
  A.~D.~Linde,
  Phys.\ Lett.\  B {\bf 259}, 38 (1991).

\bibitem{Tegmark:2005dy}
  M.~Tegmark, A.~Aguirre, M.~Rees and F.~Wilczek,
  Phys.\ Rev.\  D {\bf 73}, 023505 (2006)
  [arXiv:astro-ph/0511774].

\bibitem{Freivogel:2008qc}
  B.~Freivogel,
  arXiv:0810.0703 [hep-th].

\bibitem{Wilczek:2004cr}
  F.~Wilczek,
  arXiv:hep-ph/0408167.

\bibitem{Fox:2004kb}
  P.~Fox, A.~Pierce and S.~D.~Thomas,
  arXiv:hep-th/0409059.

\bibitem[Kamionkowski(2002)]{Kamionkowski:2002mi}
  M.~Kamionkowski,
  arXiv:astro-ph/0209273.

\bibitem[Cooray(2003)]{Cooray:2003da}
  A.~R.~Cooray,
  arXiv:astro-ph/0311059.

\bibitem{MackSteinhardt:2009}
  K.~J.~Mack and P.~J.~Steinhardt, arXiv:0911.0418 [astro-ph.CO].

\bibitem{Wantz:2009it}
  O.~Wantz and E.~P.~S.~Shellard,
  arXiv:0910.1066 [Unknown].

\bibitem{Hamann:2009yf}
  J.~Hamann, S.~Hannestad, G.~G.~Raffelt and Y.~Y.~Y.~Wong,
  arXiv:0904.0647 [hep-ph].

\bibitem{Komatsu:2008}
  E.~Komatsu et al., arXiv:0803.0547 [astro-ph].

\bibitem{Arvanitaki:2009fg}
  A.~Arvanitaki, S.~Dimopoulos, S.~Dubovsky, N.~Kaloper and J.~March-Russell,
  arXiv:0905.4720 [hep-th].

\bibitem[Kass \& Raftery(1995)]{Kass:1995}
  R.~E.~Kass \& A.~E.~Raftery,
  Journal of the American Statistical Association {\bf 90}, 773 (1995).

\bibitem{Steinhardt:2006bf}
  P.~J.~Steinhardt and N.~Turok,
  Science {\bf 312}, 1180 (2006)
  [arXiv:astro-ph/0605173].

\bibitem{Tegmark:2004qd}
  M.~Tegmark,
  JCAP {\bf 0504}, 001 (2005)
  [arXiv:astro-ph/0410281].

\bibitem{Hellerman:2005yi}
  S.~Hellerman and J.~Walcher,
  Phys.\ Rev.\  D {\bf 72}, 123520 (2005)
  [arXiv:hep-th/0508161].

\bibitem[Weinberg(1987)]{Weinberg:1987dv}
  S.~Weinberg,
  Phys.\ Rev.\ Lett.\  {\bf 59}, 2607 (1987).

\bibitem[Hall \& Nomura (2007)]{Hall:2007ja}
  L.~J.~Hall and Y.~Nomura,
  Phys.\ Rev.\  D {\bf 78}, 035001 (2008)
  [arXiv:0712.2454 [hep-ph]].

\bibitem[Linde(1994)]{Linde:1993cn}
  A.~D.~Linde,
  Phys.\ Rev.\  D {\bf 49}, 748 (1994)

\bibitem{Steinhardt:2004gk}
  P.~J.~Steinhardt and N.~Turok,
  New Astron.\ Rev.\  {\bf 49}, 43 (2005)
  [arXiv:astro-ph/0404480].


\end{thebibliography}
\end{document}